\documentclass{pasj02}
\Received{$\langle$reception date$\rangle$}
\Accepted{$\langle$acception date$\rangle$}
\Published{$\langle$publication date$\rangle$}

\usepackage[version=4]{mhchem}
\usepackage{natbib}
\usepackage[switch,mathlines]{lineno}
\usepackage{xcolor}

\begin{document}
\title{Resolved Structure and Orbital Motion of a Localized 1~au-scale Dust Accumulation in the Protoplanetary Disk around TW~Hya}
\author{
Takashi \textsc{Tsukagoshi},\altaffilmark{1}\orcid{0000-0002-6034-2892}\email{takashi.tsukagoshi.astro@gmail.com}
Takayuki \textsc{Muto},\altaffilmark{2}
Hideko \textsc{Nomura},\altaffilmark{3}\orcid{0000-0002-7058-7682}
Ryohei \textsc{Kawabe},\altaffilmark{3}\orcid{0000-0002-8049-7525}
Satoshi \textsc{Okuzumi},\altaffilmark{4}\orcid{0000-0002-1886-0880}
Shigeru \textsc{Ida},\altaffilmark{5}\orcid{0000-0001-9564-6186}
Jun \textsc{Hashimoto},\altaffilmark{6,7}\orcid{0000-0002-3053-3575}
Taichi \textsc{Uyama},\altaffilmark{7}\orcid{0000-0002-6879-3030}
Motohide \textsc{Tamura},\altaffilmark{7}\orcid{0000-0002-6510-0681}
}

\altaffiltext{1}{Faculty of Engineering, Ashikaga University, Ohmae 268-1, Ashikaga, Tochigi 326-8558, Japan}
\altaffiltext{2}{Division of Liberal Arts, Kogakuin University, 1-24-1 Nishi-Shinjuku, Shinjuku-ku, Tokyo 163-8677, Japan}
\altaffiltext{3}{Division of Science, National Astronomical Observatory of Japan, Osawa 2-21-1, Mitaka, Tokyo 181-8588, Japan}
\altaffiltext{4}{Department of Earth and Planetary Sciences, Institute of Science Tokyo, Ookayama 2-21-1, Meguro, Tokyo 152-8551, Japan}
\altaffiltext{5}{Earth-Life Science Institute, Institute of Science Tokyo, Ookayama 2-21-1, Meguro, Tokyo 152-8550, Japan}
\altaffiltext{6}{Academia Sinica Institute of Astronomy \& Astrophysics (ASIAA), 11F of Astronomy-Mathematics Building, AS/NTU, No.1, Sec. 4, Roosevelt Road, Taipei 106319, Taiwan}
\altaffiltext{7}{Astrobiology Center, National Institutes of Natural Sciences, Osawa 2-21-1, Mitaka, Tokyo 181-8588, Japan}


\maketitle

\begin{abstract}
    We present the results of high-resolution ($\sim1$~au) Atacama Large Millimeter/submillimeter Array (ALMA) observations of the TW~Hya protoplanetary disk in the Band 6 dust continuum, as well as the \ce{^13CO} and \ce{C^18O} J=2--1 emission lines.
    The primary focus of this study is to investigate the kinematics, internal morphology, and local gas environment of the prominent dust blob at a radius of 52~au.
    By comparing our 2021 data with archival observations from 2017, we detect the proper motion of the blob.
    The measured azimuthal velocity of 3.3$\pm$0.9~km~s$^{-1}$ is fully consistent with local Keplerian rotation.
    Combined with the lack of significant radial migration over the four-year baseline, this confirms that the structure is robustly co-moving with the disk system.
    Crucially, our high-resolution continuum map resolves the blob into a distinct double-peaked morphology separated by $\sim$1.7~au azimuthally.
    We robustly validate this double-peaked substructure by reproducing it in the independent 2017 dataset using a sparse-modeling image reconstruction technique.
    We discuss potential physical origins for this double-peaked morphology, including an inclined circumplanetary disk with an inner dust cavity, the roots of planet-induced spiral arms, or alternative hydrodynamic scenarios that do not involve an actively accreting planet such as the U-turn trajectory of secondary dust or a short-lived hydrodynamic gas vortex.
    We detect no compact gas emission counterparts associated with the continuum blob.
    Since these CO lines likely trace optically thick upper atmospheric layers, the absence of localized vertical gas perturbations suggests that if an embedded planet is responsible for the dust structure, its mass must be exceptionally low.
\end{abstract}


\section{Introduction}

Protoplanetary disks (PPDs) are the birthplaces of planets \citep[e.g.,][]{bib:hayashi1981}.
Direct observational evidence of actively forming planets within the PPDs around a young stars is crucial for understanding the formation and diversity of (exo-)planets.
In recent years, high-resolution observations by the Atacama Large Millimeter/submillimeter Array (ALMA) have significantly advanced our understanding of PPD structures.
Rather than being smooth and featureless, many PPDs are known to exhibit a rich variety of substructures, such as rings, deep gaps, spiral arms, or large-scale lopsided dust distribution \citep[e.g.,][]{bib:andrews2018,bib:cieza2019}.
The diversity in PPDs are believed to be a sign of ongoing planet-disk interactions, where unseen embedded protoplanets perturb the surrounding environments and cause the accumulation of dust.

An actively forming planet is expected to be surrounded by a circumplanetary disk (CPD) from which materials continuously accrete onto the planetary core \citep{bib:canup2002}.
Consequently, CPDs represent some of the promising targets for directly detecting forming planets within PPDs, as they are predicted to be observable at (sub)millimeter wavelengths \citep{bib:zhu2016,bib:szulagyi2018}.
A prominent candidate for a CPD surrounding an actively forming planet is the localized asymmetry in the dust continuum emission discovered within the large inner cavity of the PDS~70 system \citep{bib:isella2019}.
Subsequent observations revealed near-infrared scattered light and H$_\alpha$ emission counterparts, confirming the presence of actively accreting protoplanets at the positions of PDS 70b and 70c \citep{bib:keppler2018,bib:haffert2019}.
However, while planets are generally thought to be formed deeply embedded within their parent PPDs, the observational footprints of candidate CPD in such embedded environments have yet to be reported except for a small-scale dust blob around TW~Hya at a radius of 52~au \citep{bib:tsukagoshi2019a}.

TW~Hya is one of the most studied young stellar objects hosting a PPD.
Located at a distance of 60~pc \citep{bib:bailer2018}, it represents the closest gas-rich PPD to Earth.
Its face-on geometry, with an inclination of 5$\degree$ \citep{bib:bailer2018}, makes it an ideal laboratory for spatially resolving fine disk substructures without projection effects.
While previous high-resolution observations of TW~Hya have successfully identified multiple ring and gap structures across the dust disk \citep{bib:andrews2016,bib:tsukagoshi2016}, the nature of the compact, non-axisymmetric dust blob at a radius of 52~au discovered by \citet{bib:tsukagoshi2019a} remains an active topic of debate.
They pointed out that this non-axisymmetric structure could trace a dust-rich CPD around an actively accreting Neptune-mass planet, or alternatively, an accumulation of dust trapped within a small-scale gas vortex driven by disk instabilities.
More recently, detailed hydrodynamic simulation by \citet{bib:zhu2023} demonstrated that the blob could correspond to a dense envelope and planet-induced spiral arms surrounding a forming planet.
Conversely, alternative scenarios that do not involve an actively forming planet have also been proposed, such as the localized accumulation of secondary dust grains moving along a "U-turn" trajectory after being ejected from a migrating planetary core \citep{bib:nayakshin2020}.

In this paper, we present high-resolution ($\sim1$~au) ALMA Cycle 7 observations of the TW~Hya protoplanetary disk in the Band 6 dust continuum, as well as the \ce{^13CO} and \ce{C^18O} J=2--1 emission lines.
The aims of our observations are to definitely measure the kinematics of the 52~au blob, reveal its internal substructures, and constrain its local gas environment.
The paper is organized as follows.
Section \ref{sec:observations} describes the ALMA observations and data reduction procedures.
In Section \ref{sec:results}, we present the results of the continuum and molecular line imaging, including the detection of the proper motion of the blob and its double-peaked substructure.
In Section \ref{sec:discussion}, we discuss the potential physical origins of the blob and the implications of the molecular gas non-detection.
Finally, our main findings and conclusions are summarized in Section \ref{sec:summary}.

\section{Observations and Imaging}\label{sec:observations}

\subsection{Observations}
We conducted our ALMA Band 6 observations of TW~Hya in the long-baseline configuration (C-10) during Cycle 6 and 7 periods [Project ID: 2018.1.01173.S]. 
The observations consisted of six execution blocks (EBs) in total. 
One EB was executed in July 2019, while the remaining five were executed in August and September 2021. 
The total on-source integration time was 4.7 hours. 
Three spectral windows (SPWs) were configured in the Time Division Mode (TDM) to detect the Band~6 continuum emission, each with a total bandwidth of 1.875~GHz. 
The remaining SPW was configured in the Frequency Division Mode (FDM) to simultaneously observe the \ce{^13CO} and \ce{C^18O} J=2--1 lines. 
This SPW had a total bandwidth of 937.5~MHz and a channel width of 488.281~kHz, corresponding to a velocity resolution of 1.33~km~s$^{-1}$.

Data reduction and initial calibrations for bandpass characteristics and complex gain fluctuations, as well as all subsequent imaging processes for these newly obtained data, were carried out using the Common Astronomy Software Applications (CASA) package version 6.4.0.

\subsection{Dust Continuum Calibration and Imaging}\label{sec:cont_calib}
Before combining the calibrated visibilities from all EBs, we corrected for the proper motion of the target, using the July 2019 EB as the reference frame. 
We generated a preliminary dirty image for each EB and determined the central emission peak of the disk by performing a 2D Gaussian fit using the CASA task {\it imfit}. 
The phase center of each EB was then shifted to this determined emission center using the task {\it fixvis}. 
Finally, all the phase-aligned datasets were combined using the {\it concat} task, applying a tolerance parameter sufficient to merge the pointing centers.
Because the subsequent phase self-calibration process modifies the complex gains, absolute coordinate information is inherently lost.
Consequently, all resulting combined and self-calibrated images are in a relative coordinate system centered on the continuum peak of the disk.

To obtain a high-fidelity continuum image, we performed iterative self-calibration
on the concatenated visibilities. Imaging during this process was performed using
the {\it tclean} task with Briggs weighting and a robustness parameter of 0.5.
We employed the multi-scale deconvolver with scale parameters of 0 (point source),
1, 3, 5, 8, and 10 times the synthesized beam size. The CLEAN masking was
handled automatically using the {\it auto-multithresh} algorithm. The self-calibration
process consisted of five rounds of phase-only calibration followed by one round
of amplitude calibration. In each iteration, the resulting CLEAN model was used
to solve for the gain solutions of the subsequent step. For the phase-only
calibration, the solution intervals were progressively decreased: '{\it inf}'
(the entire observation time of an EB), 1200, 600, 300 and 120 sec. The
final amplitude calibration was performed with a solution interval of '{\it inf}'.

Following the self-calibration derived from the full combined dataset, we generated
the final CLEAN map of the continuum emission using only the 2021 data. This
approach was adopted to mitigate any potential blurring effects caused by
the time variation (i.e., proper motion) of the substructures in the disk. The
resulting synthesized beam size of this 2021 continuum image is 22.0$\times$19.9~mas
with a position angle (PA) of $-62\fdg9$, which corresponds to a spatial
resolution of 1.3$\times$1.2~au at an assumed distance of 60~pc. The self-calibration
significantly improved the image quality, achieving an rms noise level of
5.3 $\mu$Jy beam$^{-1}$.

The maximum recoverable scale (MRS) of the C-10 configuration was approximately $0\farcs22$~arcsec. 
This implies that a significant fraction of the extended disk emission (roughly 32\% of the total flux density) is filtered out. 
To recover the continuum emission from the entire disk, we concatenated our long baseline data with archival Band 6 data taken with shorter baselines.
Specifically, we used data from Cycle 4 (Project ID: 2016.1.00842.S; observed in 2017 with the C40-7 configuration), which was previously calibrated and published by \citet{bib:tsukagoshi2019a}. 
The MRS of this short-baseline dataset is approximately $2\farcs6$, sufficient to capture the entire dust disk. 
We combined this archival dataset with our self-calibrated long-baseline data. 
To align the emission between the two epochs, we performed one round of phase-only self-calibration
on the combined visibilities using a solution interval of '{\it inf}'. 
The combined dataset was then imaged using the same {\it tclean} parameters as the long-baseline data, except for an adjusted noise threshold. 
The resulting synthesized beam size of this combined continuum map was $21.5\times19.7$~mas with a PA of $-66\fdg4$, with an rms noise level of 5.8~$\mu$Jy~beam$^{-1}$.

\subsection{Molecular Line Calibration and Imaging}\label{sec:line_calib}
For our long-baseline molecular line data, we first performed continuum subtraction in the visibility domain.
Subsequently, the gain solutions derived from the continuum self-calibration of our data were applied to these line measurement sets to perform the necessary phase and amplitude corrections.

To recover the extended molecular gas emission and properly evaluate the gas environment around the blob, we also incorporated existing short-baseline archival data for the \ce{^13CO} and \ce{C^18O} lines [Project IDs: 2016.1.01375.S and 2016.1.00229.S].
The baseline lengths range from 15 to 700~m, which effectively fill the missing short uv coverage of our observations.

Prior to combining the molecular line datasets, the archival data were independently processed.
For these archival datasets, the initial data calibration was performed using the standard ALMA pipeline scripts with CASA version 4.7.2, while the subsequent imaging processes utilized version 6.7.0.
For each calibrated dataset, we performed an iterative CLEAN imaging with self-calibration on the continuum spectral windows of the archival data to generate calibration tables. 
Following continuum subtraction, these tables were applied to the respective archival line measurement sets.
We applied the same phase-center shifting and concatenation described in \S\ref{sec:cont_calib} to the \ce{^13CO} and \ce{C^18O} line data.
The final gain solutions derived from the combined continuum self-calibration were then applied to these fully concatenated line datasets.

For the final line imaging, we created data cubes consisting of 50 velocity channels, starting from $-15$~km~s$^{-1}$ with a velocity width of 0.67~km~s$^{-1}$. 
The imaging was performed using the multi-scale CLEAN with Briggs weighting.
The robust parameter was set to 0 to suppress the broad wings of the point spread function introduced by the short baselines.
An uvtaper of $0\farcs1$ was also applied to improve the image sensitivity.
The resulting synthesized beam size and rms noise level per channel are $0\farcs13\times0\farcs12$ (PA$\sim-47\deg$) and 0.55 mJy~beam$^{-1}$ for both the \ce{^13CO} and \ce{C^18O} lines.

\section{Results}\label{sec:results}

\subsection{Continuum map}
\begin{figure*}[htb]
    \begin{center}
        \includegraphics[width=80mm]{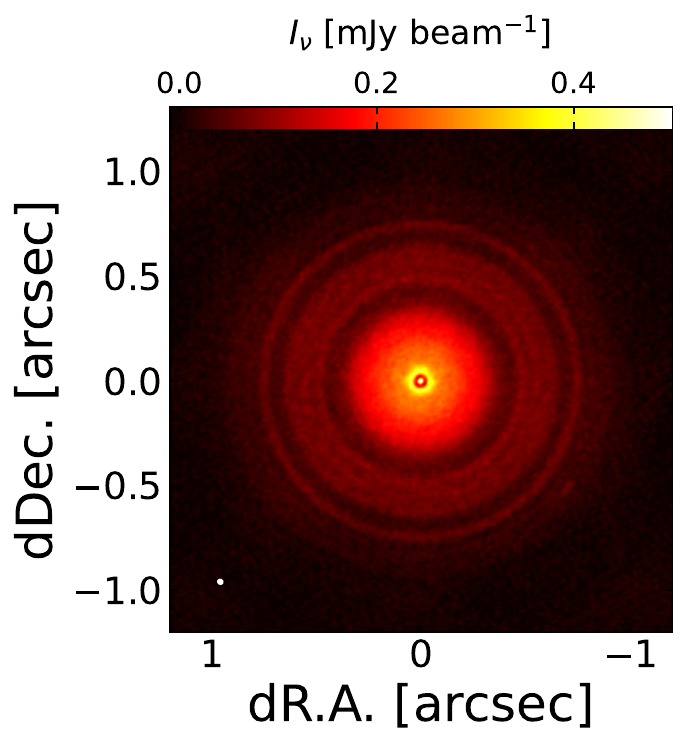}
        \includegraphics[width=80mm]{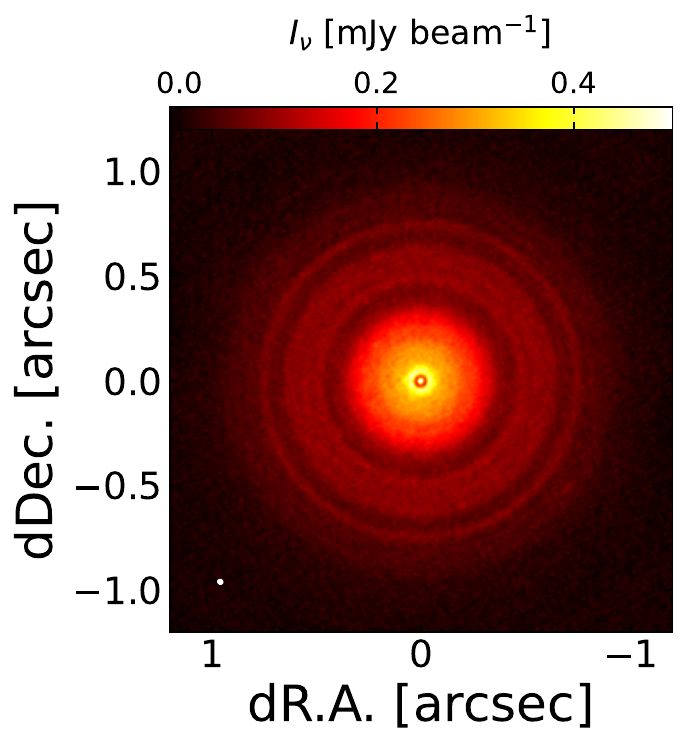}
    \end{center}
    \caption{(Left) Band6 continuum map reconstructed from our 2021 data
        only. The circle in white at the bottom right corner indicates the beam size.
        (Right) Band6 continuum map reconstructed from the combined data where
        our longest baseline data and the archival short baseline data are
        concatenated. {Alt text: Two side by side intensity maps of a protoplanetary disk. Axes show positional offsets from the disk center in arcseconds. The intensity unit is millijanskys per beam. The both maps show resolved concentric rings, with the right map showing smoother extended structure.}}
    \label{fig:cont_map}
\end{figure*}

\begin{figure}
    \begin{center}
        \includegraphics[width=80mm]{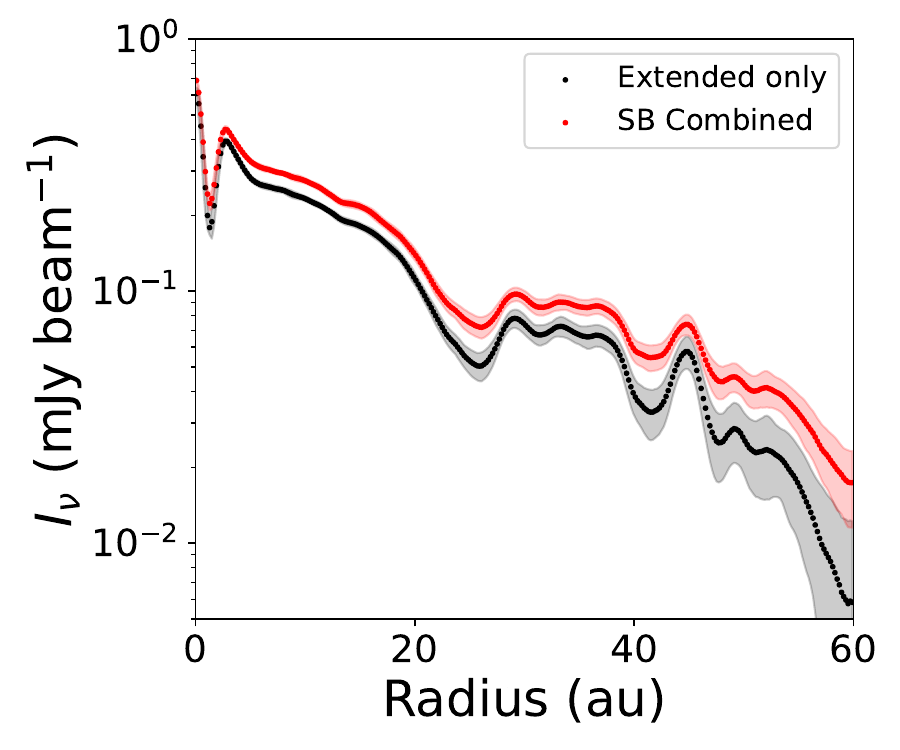}
    \end{center}
    \caption{Radial profiles of the continuum map averaged over the entire
        position angle. The profile obtained from the longest baseline data only
        is shown by black points, and the shaded area indicates the standard
        error through the averaging. The profile of the combined map of the long
        and short baseline data is shown in red. {Alt text: A line graph of the radial profiles of continuum emission.The horizontal axis is radius in astronomical units, and the vertical axis is intensity in millijanskys per beam. Two plotted curves exhibit multiple local peaks, with one showing consistently higher intensity.}}
    \label{fig:cont_profile}
\end{figure}

The high-resolution Band\ 6 continuum images of the TW~Hya protoplanetary disk are presented in Figure\ \ref{fig:cont_map}.
The left panel of Figure\ \ref{fig:cont_map} displays the image reconstructed exclusively from our 2021 long-baseline observations.
While this map is highly sensitive to compact structures, a significant portion of the extended disk emission is resolved out.
The right panel of Figure\ \ref{fig:cont_map} shows the image combined with the archival short-baseline data, which successfully recovers the entire disk structure and the total flux density.
The total flux density measured from the longest baseline image is 371~mJy, whereas the combined image yields a total flux density of 544~mJy.
This recovered value is fully consistent with previous observational studies \citep{bib:tsukagoshi2016,bib:huang2018}.
In both images, the overall axisymmetric ring/gap structures characteristic of this system are clearly resolved.
Figure\ \ref{fig:cont_profile} presents the azimuthally averaged radial profiles of the continuum emission derived from the images in Figure\ \ref{fig:cont_map}, quantitatively confirming the presence and locations of these multiple annular structures.
The combination of short-baseline data effectively restores the smooth background emission of the entire disk, whereas the 2021-only data intrinsically acts as a spatial filter, optimally highlighting localized, small-scale structures.
With the improved spatial resolution and sensitivity, our new maps reveal several updated substructures compared to previous observations \citep{bib:andrews2016,bib:tsukagoshi2019a}.
Specifically, we clearly detect the compact inner disk at $r<1$~au, whose flux density is estimated in the combined image to be 1.18$\pm$0.01~mJy.
Furthermore, the innermost gap structure immediately outside this inner disk is distinctly resolved.
In the outer disk, the prominent gap at 25~au is now resolved into finer components, and delicate substructures are clearly visible within the emission plateau around 30~au.

\begin{figure}[htb]
    \begin{center}
        \includegraphics[width=80mm]{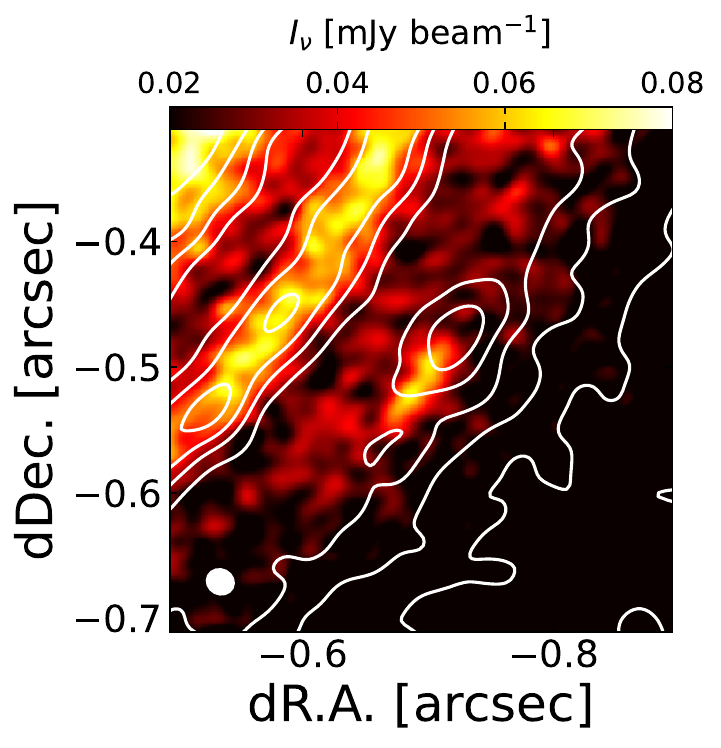}
    \end{center}
    \caption{Positional variation of the mm blob over time from 2017 to
        2021. The color scale shows the continuum map reconstructed from our
        longest baseline data only. The white contour indicates the continuum map
        obtained in \citet{bib:tsukagoshi2019a}. The contour interval is 5$\sigma$,
        where 1$\sigma$ is 9.1~$\mu$Jy~beam$^{-1}$. {Alt text: A close up intensity map comparing two observation epochs. Axes display positional offsets in arcseconds. An underlying map is overlaid with contour lines, showing their respective centers slightly offset from each other, with the contour line slightly shifted counter-clockwise from the underlying map.}}
    \label{fig:blob_closeup}
\end{figure}

Focusing on the most prominent non-axisymmetric feature, Figure\ \ref{fig:blob_closeup} provides an enlarged view of the localized, blob-like continuum emission located at a radial distance of approximately 52~au in the south-east region of the disk, which was initially identified by \citet{bib:tsukagoshi2019a}.
In this figure, the continuum image obtained from our 2021 observations is presented in color, overlaid with white contours representing the Cycle\ 4 data from \citet{bib:tsukagoshi2019a}.
A visual comparison clearly reveals that the blob has moved over time, shifting toward the south-east direction relative to its previous position in the 2017 epoch.
To quantify this positional shift, we derived the exact locations of the blob in both epochs.
For a consistent comparison, our 2021 continuum map was smoothed to match the angular resolution of the \citet{bib:tsukagoshi2019a} image.

Because the absolute coordinate information is lost after self-calibration, we measured the proper motion of the blob in a relative coordinate frame.
We performed a 2D Gaussian fit to determine the exact position of the blob relative to the central continuum peak of the disk in each respective epoch.
The fitting precision of the disk center is approximately 1~mas, which is sufficiently smaller than the positional fitting error of the blob itself, ensuring a highly precise relative comparison between the two epochs.
We then performed a 2D Gaussian fit to the blob emission in each image.
The fitting results yielded a radial distance of $r=0\farcs863\pm0\farcs010$ and PA of $233\fdg95\pm0\fdg74$ for our 2021 data, while the 2017 data yielded $r=0\farcs862\pm0\farcs023$ and PA of $236\fdg88\pm2\fdg36$ with respect to the disk center.
The negligible difference in the radial positions indicates that no significant radial motion of the blob occurred over the four-year observation period.
The measured offset between the two epochs is 44.2$\pm$12.1~mas to the south-east, which corresponds to a physical distance of 2.7$\pm$0.7~au.
Given the roughly four-year duration between the observations, we estimate the proper motion velocity of the blob to be 3.3$\pm$0.9~km~s$^{-1}$.
This value is consistent with the expected Keplerian velocity of $\sim$3.4~km~s$^{-1}$ at a radius of 52~au in the TW~Hya disk.
This excellent agreement strongly indicates that the blob-like structure is co-moving with the Keplerian rotation of the disk system.

We estimated the flux density of the blob using the map combined with the short-baseline data.
We measured the total flux density within the blob's emission region (approximately $4\times1$~au) as identified by \citet{bib:tsukagoshi2019a}, and subtracted the local background emission from the surrounding protoplanetary disk.
This direct image-based approach yields an excess flux density of 193$\pm$25~$\mu$Jy for the blob.
This derived value is slightly smaller than the estimate of 250~$\mu$Jy reported by \citet{bib:tsukagoshi2019a}.
A plausible reason for this discrepancy is a potential overestimation in the previous study.
\citet{bib:tsukagoshi2019a} evaluated the blob flux after subtracting an axisymmetric disk model, however, residual systematic errors from this subtraction might have artificially enhanced the flux density at the location of the blob.
Furthermore, their residual map after performing a Gaussian fit to the blob exhibits a localized distribution of negative components (see Fig.~3 of \citealt{bib:tsukagoshi2019a}), which indicates a possible overestimation of the flux density.
In fact, when we apply the same image-based derivation to the image from \citet{bib:tsukagoshi2019a}, we obtain a flux density of $\sim$189~$\mu$Jy, which is consistent with the estimate in this study.

While the blob appeared as a single-peaked structure in the previous study \citep{bib:tsukagoshi2019a}, our new high-resolution observation resolves it into two distinct peaks.
The spatial separation between these two peaks is approximately $0\farcs028$, corresponding to 1.7~au at the distance of TW~Hya.
In the combined map, the absolute peak intensities before background subtraction are 74.1 and 73.3~$\mu$Jy~beam$^{-1}$ for the northern and southern components, respectively.
To quantify the excess emission of the blob, we evaluated the local background emission from the underlying protoplanetary disk to be 39.6~$\mu$Jy~beam$^{-1}$ by averaging the intensities in the adjacent azimuthal regions (P.A.=$222\degree$--$252\degree$ excluding the region of the blob), following the same procedure as \citet{bib:tsukagoshi2019a}.
Subtracting this background disk emission yields the excess peak intensities of 34.5 and 33.7~$\mu$Jy~beam$^{-1}$ for the northern and southern components, respectively.

\begin{figure}[htb]
    \begin{center}
        \includegraphics[width=80mm]{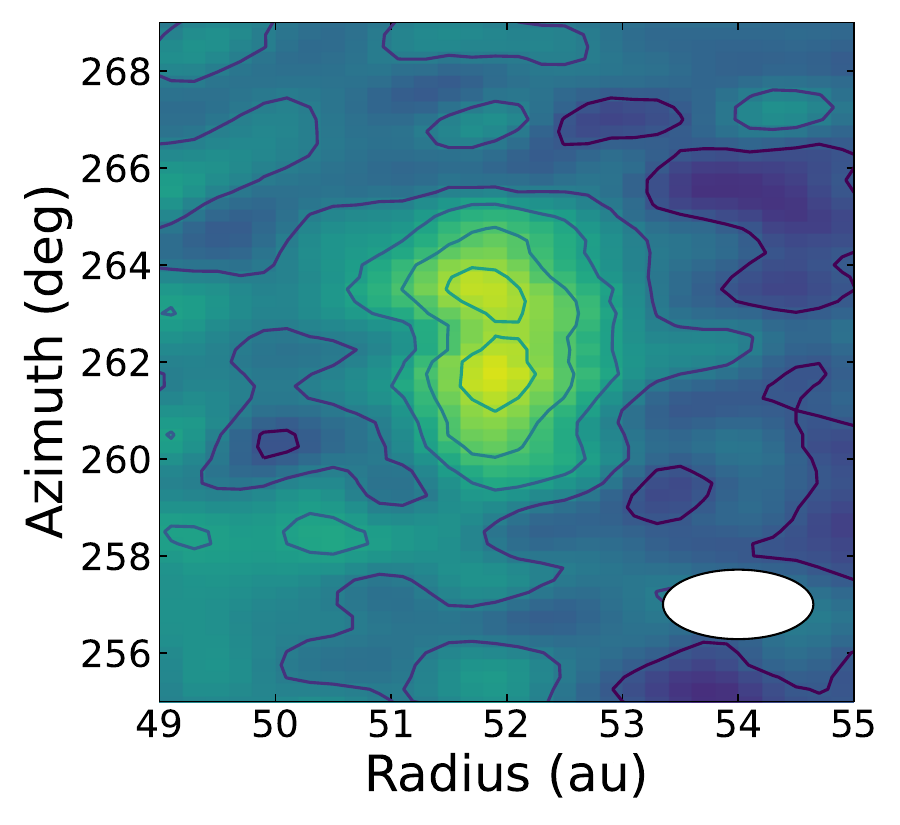}
    \end{center}
    \caption{Close-up view of the deprojected polar map around the blob.
        The white ellipse at the bottom-right corner represents the beam size.
        The contour levels are 0.02, 0.03, 0.04, 0.05, and 0.06 mJy~beam$^{-1}$. {Alt text: A deprojected polar map of a localized dust accumulation in a protoplanetary disk. The horizontal axis is radius in astronomical units, and the vertical axis is azimuth in degrees. A central high intensity region is separated into two distinct vertical peaks.}}
    \label{fig:blob_polar}
\end{figure}

Figure~\ref{fig:blob_polar} displays the deprojected polar map around the blob created from the longest baseline image.
The distribution of the emission is slightly skewed radially inward toward the upstream side of the Keplerian rotation.
We caution, however, that the radial amplitude of this skewness is significantly smaller than the synthesized beam size.
Therefore, while intriguing, this morphological feature remains tentative and requires further confirmation through future observations.

\subsection{Molecular lines}
\begin{figure*}
    \begin{center}
        \includegraphics[width=80mm]{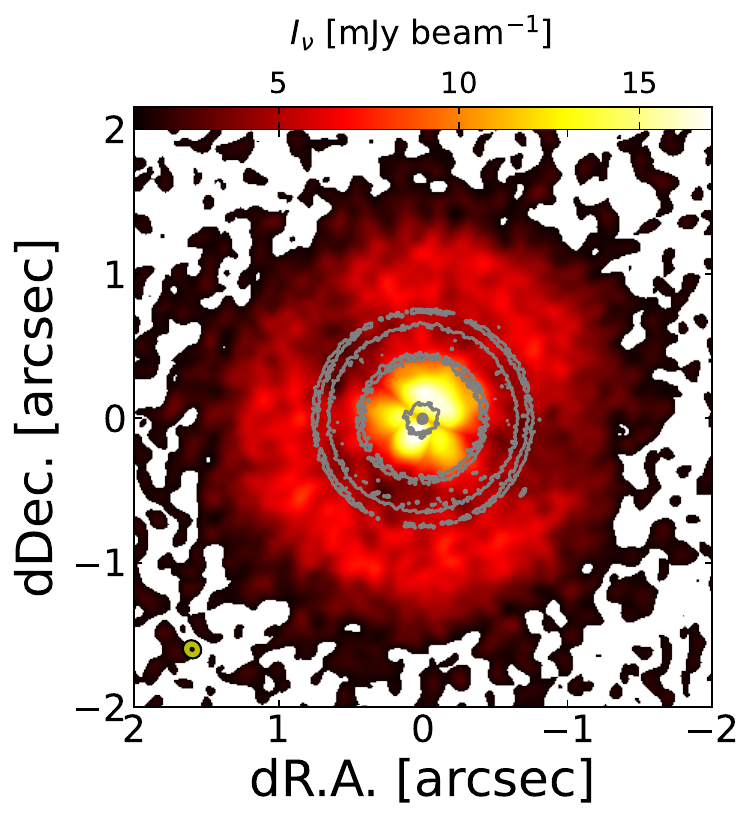}
        \includegraphics[width=80mm]{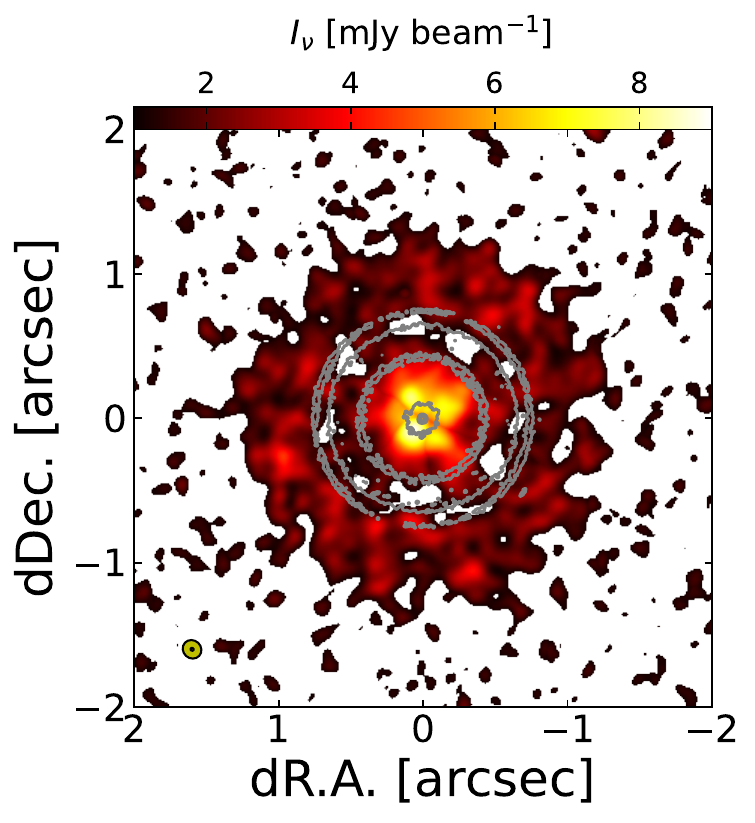}        
    \end{center}
    \caption{(Left) Peak intensity map in \ce{^13CO} reconstructed from the combined short- and long-baseline data is shown in color. The contour lines indicate 10 and 50$\sigma$ of our Band 6 continuum map reconstructed from the longest baseline data only, where 1$\sigma$ is 5.3~$\mu$Jy~beam$^{-1}$. The yellow ellipse at the bottom left corner indicates the beam size of the \ce{^13}CO map, and the gray ellipse inside it shows the beam size of the continuum map. (Right) Same as the left panel, but for the \ce{C^18O} map. {Alt text: Two side by side maps of peak gas intensity in \ce{^13}CO and \ce{C^18O} overlaid with contours of a dust continuum. Axes show positional offsets from the disk center in arcseconds. Contour lines are widely distributed with concentrating around the underlying bright central region.}}
    \label{fig:lines}
\end{figure*}

We first checked the channel maps of the \ce{^13CO} and \ce{C^18O} lines in the combined short- and long-baseline data, but found no distinct localized emission associated with the continuum blob at 52~au.
To better visualize the spatial distribution of the extended line emission and to confirm the absence of a compact peak at the blob's location, we created peak intensity (moment~8) maps for both molecular lines.
Figure~\ref{fig:lines} displays the color maps of the \ce{^13CO} and \ce{C^18O} peak intensities overlaid with contours of the Band 6 continuum emission.
Both maps reveal extended gas structures, with the \ce{^13CO} emission appearing more extended than the \ce{C^18O}, likely due to the lower optical depth of \ce{C^18O}.
Consistent with the channel maps, these images show no compact gas emission counterparts associated with the continuum blob.
To estimate the upper limits on the molecular column densities, we integrated the emission of both molecular lines over the entire velocity range where emission is detected, i.e., from 0.41 to 5.10 km~s$^{-1}$ for \ce{^13CO} and from 1.08 to 5.10 km~s$^{-1}$ for \ce{C^18O}, respectively.
From the integrated intensity maps, we estimate 3$\sigma$ upper limit of 4.8 and 4.3 mJy~beam$^{-1}$~km~s$^{-1}$ for \ce{^13CO} and \ce{C^18O} at the location of the blob, respectively.
Assuming local thermodynamic equilibrium (LTE) and an excitation temperature of 30~K, this upper limit yields bounds on the molecular column densities of $N$(\ce{^13CO}) $< 4.0 \times 10^{17}$~cm$^{-2}$ and $N$(\ce{C^18O}) $< 2.7 \times 10^{18}$~cm$^{-2}$, under the assumption of optically thin emission.
Because CO line emission generally becomes optically thick in the upper layers of the disk atmosphere \citep{bib:zhang2017,bib:nomura2021}, these non-detections may not accurately reflect the underlying gas density structure near the disk midplane where the dust blob resides.

\section{Discussion}\label{sec:discussion}

\subsection{Confirmation of the double-peaked substructure in the blob}
To confirm this double-peaked substructure, we applied an image reconstruction technique based on sparse modeling to the independent archival dataset from \citet{bib:tsukagoshi2019a}.
For this purpose, we utilized PRIISM \citep{bib:priism}, an imaging tool optimized for high-resolution imaging of ALMA continuum data, in which the underdetermined observation equation is solved by employing two penalty terms: the $L_{1}$ norm and the Total Squared Variation (TSV).
By providing a higher spatial resolution than conventional CLEAN imaging, this tool has been successfully applied to resolve substructures in various protoplanetary disks \citep{bib:yamaguchi2020,bib:yamaguchi2024,bib:shoshi2025}.

\begin{figure}[htb]
    \begin{center}
        \includegraphics[width=80mm]{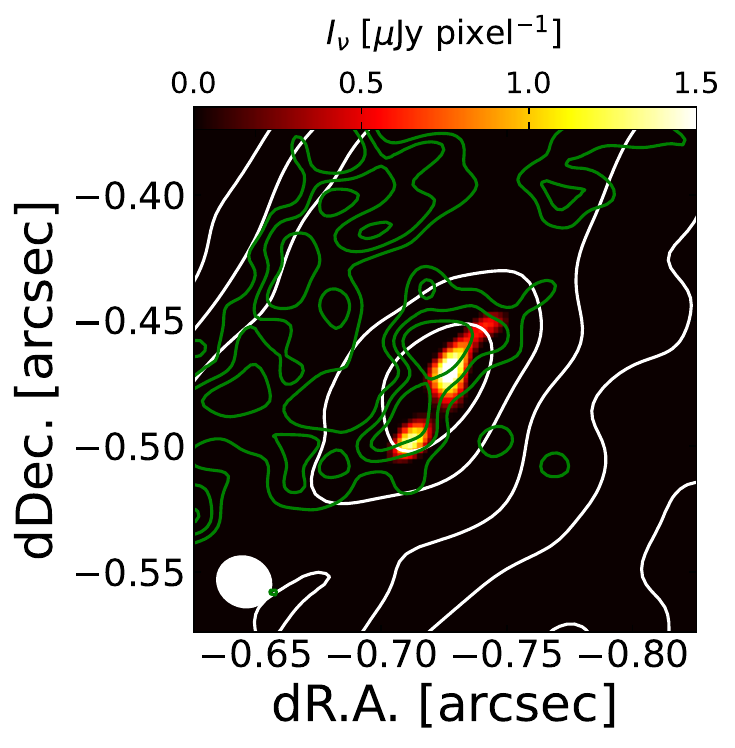}
    \end{center}
    \caption{Higher-resolution image of the 2017 data \citep{bib:tsukagoshi2019a}
        reconstructed with PRIISM. The PRIISM image is shown in color, while the
        white contour represents the original CLEAN image. The contour interval
        is the same as in Figure~\ref{fig:blob_closeup}. The white ellipse at the
        bottom left corner indicates the beamsize of the original CLEAN image.
        The green contour shows the continuum map of our 2021 data, after applying
        a 2$\fdg$93 rotation with respect to the disk center to match the position
        angle to the 2017 data. The contours represent 9, 10, and 11$\sigma$. {Alt text: A close up view of the resolved intensity map comparing two datasets. Axes represent spatial coordinates in arcseconds. A central bright region splits into two peaks in azimuth.}}
    \label{fig:priism}
\end{figure}

We applied PRIISM to the residual visibilities of the 2017 data, from which the axisymmetric disk emission component had been subtracted.
We employed the FFT version of the image reconstruction because the massive number of visibilities in our dataset rendered the non-uniform FFT approach computationally prohibitive.
The image was reconstructed using a 1024$\times$1024 pixel grid with a cell size of 0.002 arcsec.
The optimal hyperparameters for the penalty terms were determined through a 10-fold cross-validation process.
To avoid overfitting and the amplification of spurious noise, we selected the optimal image by evaluating both the mean squared error (MSE) and the overall image fidelity.
Figure~\ref{fig:priism} shows the resulting PRIISM image of the blob derived from the 2017 data, which successfully recovers the two distinct emission peaks within the blob.
This consistent detection in an independent dataset strongly validates that the double peaked morphology is real rather than an imaging artifact.

Figure~\ref{fig:priism} also reveals a marginal inward radial offset in the position of the blob between the two epochs.
However, as noted in Section~\ref{sec:results}, this radial offset is smaller than the synthesized beam size.
Given the potential systematic uncertainties introduced by the image reconstruction techniques and the precise derivation of the disk center, we conservatively treat this radial offset as negligible in the current study.
Further confirmation through future higher-resolution observations is required to verify this radial migration.
Nevertheless, if this tentative inward shift reflects a physical migration of the blob, it would correspond to a rapid radial drift velocity of approximately 70~m~s$^{-1}$.
Such a rapid inward drift implies an extremely short dynamical timescale, suggesting that the dust structure would migrate into the central star in merely 3500~years.

\subsection{Origin of the millimeter blob}
The fundamental nature of the small-scale dust blob at 52~au has been a subject of debate since its initial discovery \citep{bib:tsukagoshi2019a}.
The estimated dust mass of the blob \citep[$\sim0.03$~$M_{\oplus}$; ][]{bib:tsukagoshi2019a,bib:ilee2022} and the lack of a deep, fully carved annular gap place strong constraints on the mass of any embedded planet, restricting it to the mass regime of a Neptune or a super-Earth.
Our new high-resolution images have resolved this localized continuum emission into a clear double-peaked morphology, with the two components separated by approximately 1.7~au.
This observed substructure provides observational constraints for theoretical models of disk-planet interactions    and instabilities.
In this section, we discuss potential physical mechanisms that could potentially produce the observed double-peaked morphology.

\subsubsection{Inclined circumplanetary disk with an inner hole}
If a forming planet is currently accreting gas and dust at 52~au, it is expected to host a circumplanetary disk (CPD) with a characteristic size of a fraction of its Hill radius \citep[e.g.,][]{bib:canup2002,bib:wang2014}.
The $\sim1.7$~au separation between the two peaks observed in our data is highly consistent with the expected spatial scale of a CPD for an ice giant, as previously suggested by \citet{bib:tsukagoshi2019a}.
A natural explanation for the double-peaked morphology in this scenario is that the CPD itself contains an inner cavity or a ring-like dusty ring.
Indeed, such a ring-like CPD geometry has recently been proposed for the CPD candidate around PDS~70c, where an optically thick dust ring is favored to consistently explain its multi-wavelength continuum emissions \citep{bib:shibaike2026}.

Theoretical calculations suggest that the size of a CPD is approximately 1/10 to 1/3 of the Hill radius \citep[e.g.,][]{bib:quillen1998,bib:wang2014}.
Assuming that the separation between the two observed peaks corresponds to the diameter of a CPD or a dusty ring, the radius would be 0.85~au.
If this radius roughly represents the outer extent of the CPD, it would correspond to 1/3 of the Hill radius for a planet with 5 Neptune masses, or 1/10 of the Hill radius for a 10 Jupiter-mass planet.
However, a massive gas giant in the latter case would dynamically carve a deep and wide gap in the background protoplanetary disk \citep[e.g.,][]{bib:lin1993}.
Given the lack of such a prominent gap structure at 52~au, the lower-mass case is considered much more preferable.

If a dusty CPD forms a ring-like structure that is significantly inclined relative to the plane of the sky, i.e., misaligned with the main TW~Hya's PPD, the projection of the optically thin ring onto our line of sight would result in strong limb brightening along its major axis.
This geometric effect naturally produces two distinct emission peaks when observed at a finite spatial resolution, as seen in observations of ring-like structures of protoplanetary disks at moderate spatial resolutions \citep[e.g.,][]{bib:andrews2011,bib:tsukagoshi2014}.
It is worth noting that such a misalignment between the rotation axes of the CPD and the PPD is theoretically possible even during the primordial embedded phase.
For instance, global magnetohydrodynamic simulations have demonstrated that stochastic accretion flows into the Hill sphere can generate an inclined CPD \citep{bib:gressel2013}.
Furthermore, a recent study suggests that such tilted CPDs can grow and survive under certain physical conditions \citep{bib:martin2020}.

Using the excess peak intensities derived in \S~\ref{sec:results} (34.5 and 33.7~$\mu$Jy~beam$^{-1}$ for the north and south peaks, respectively), we can estimate the average dust column density of this inclined CPD or dusty ring within our synthesized beam under the assumption of optically thin emission.
Here, we adopt the dust temperature at the disk midplane of 18~K, based on previous modeling studies of the submillimeter continuum emission from the TW~Hya disk \citep{bib:zhang2017,bib:huang2018}.
The dust mass opacity coefficient is assumed to be 2.3~cm${^2}$~g$^{-1}$ \citep{bib:tsukagoshi2019a}.
With these assumptions, the line-of-sight column densities, $\Sigma_\mathrm{CPD}/\cos \theta$, are estimated to be 0.06~g~cm$^{-2}$ for both the northern and southern peaks, where $\theta$ is the inclination angle of the CPD relative to the line of sight.
However, we caution that if the dust emission is not fully optically thin, subtracting the background disk emission leads to an underestimation of the intrinsic dust mass of the clumps.
Therefore, the column densities derived here from the excess flux should be regarded as lower limits.

It should be noted that our revised estimate still indicates a substantial total flux density of $\sim$200~$\mu$Jy for the entire blob structure (see \S~\ref{sec:results}).
As pointed out by \citet{bib:tsukagoshi2019a}, it is theoretically challenging to account for such a high flux density solely with a CPD.
Therefore, to fully explain the observed emission within this scenario, it would likely be necessary to consider the presence of an additional, surrounding envelope-like structure.

\subsubsection{Roots of planet-induced spiral arms}
An alternative planet-induced scenario involves spiral structures induced by an embedded planet.
Theoretical calculations suggest that a planet with a mass lower than the thermal mass of the parent PPD cannot carve a clear, observable gap, nor can it form a distinct CPD.
Instead, a dense envelope would be formed around the planet, which could be an observable localized structure \citep{bib:zhu2023}.
The expected mass for the planet associated with the 52~au blob is in the Neptune regime \citep{bib:tsukagoshi2019a}, which matches this theoretical scenario.

According to the simulations by \citet{bib:zhu2023}, if the background protoplanetary disk is optically thin enough to allow the detection of the optically thick localized blob embedded in the disk midplane, both the planetary envelope and the spiral structure induced by the planet should be visible.
Furthermore, their models predict that when the background disk is in a marginally optically thick regime, this excited spiral structure within the envelope can be observed as a spatially extended feature that appears as two distinct peaks.

Current observational estimates for the TW~Hya disk are consistent with this condition.
For the background disk, \citet{bib:tsukagoshi2022} estimated an optical depth of $\tau\sim0.2$ at 52~au. Even when considering the effects of dust scattering, the optical depth is constrained to be $\tau\sim2$ at a wavelength of 0.9~mm \citep{bib:macias2021}.
Therefore, the background disk at the location of the blob is expected to be optically thin to marginally optically thick. 
Evaluating the spectral index of the blob using the Band~6 flux density derived in this study and the previous Band~7 observation \citep{bib:ilee2022} yields $\alpha=3.4\pm1.5$.
A natural explanation of this value is that the blob cannot be characterized as optically thick.
However, due to the close frequency proximity between Band~6 and Band~7, a large uncertainty still persists, leaving an optically thick scenario as a plausible solution.
Future high-resolution and high-sensitivity observations at lower frequencies will be necessary to accurately isolate the blob's emission and constrain its optical depth.

Regarding the spatial scale of such features, theoretical simulations predict that the extended structures excited around a Neptune-mass protoplanetary core can reach a size comparable to the local gas pressure scale height, particularly when the parent PPD is marginally optically thick.
Assuming a local disk temperature of 18~K at a radius of 52~au \citep{bib:zhang2017,bib:huang2018}, the scale height is estimated to be approximately 4~au \citep{bib:tsukagoshi2019a}.
While the observed separation of 1.7~au between the two peaks is bit smaller than this nominal scale height, this morphology might still be well explained by considering specific local conditions.
For instance, a localized decrement in temperature would naturally reduce the local scale height.
Combined with potential variations in the local disk opacity at the position of the blob, these factors could reasonably account for the observed spatial separation within this scenario.

\subsubsection{Alternative scenarios without an actively forming planet}
While the scenarios discussed above rely on the presence of a growing protoplanet actively accreting surrounding gas and dust, alternative mechanisms can also produce the observed double-peaked substructure without requiring an actively forming planet at the center of the blob.

One such possibility involves the dynamics of secondary dust particles released from an already grown planetary core.
According to the model proposed by \citet{bib:nayakshin2020}, this process can naturally produce an azimuthally elongated dust structure through the interaction between the released dust and the surrounding gas.
In this scenario, the dust grains are initially small and thus strongly coupled to the gas.
Because the gas at the planet's orbit rotates at a slightly sub-Keplerian velocity due to the global radial pressure gradient, these small dust particles begin to fall behind the planet.
Subsequently, the dust grains undergo rapid size growth and start to drift radially inward due to aerodynamic gas drag.
As they migrate to inner orbits, their angular velocity increases.
Eventually, their angular velocity exceeds that of the planet, causing the dust particles to catch up with and overtake the planet.
This kinematic sequence creates a distinct "U-turn" trajectory.
Within this framework, an intermittent or episodic ejection of secondary dust from the core could introduce spatial density variations along the trajectory.
Such inhomogeneities could naturally lead to localized dust accumulations, potentially yielding the two distinct peaks observed in our high-resolution map.

Another scenario that explains the observed double-peaked morphology is a small gas vortex generated by disk instabilities.
As pointed out by \citet{bib:tsukagoshi2019a}, the accumulation of dust trapped within the small gas vortex could be responsible for the observed blob.
If the disk is turbulent, such instabilities would be expected to form multiple gas vortices, potentially resulting in several dust blobs distributed along the same orbital radius \citep[e.g.,][]{bib:raettig2015,bib:ono2018}.
However, even with the high sensitivity of our observations, we did not detect any additional small-scale dust blobs along the 52~au orbit.

\citet{bib:richard2016} demonstrated that when the local cooling time of the disk gas is short, small and short-lived vortices can be formed.
In the outer regions of a protoplanetary disk, where the radial temperature and density profiles tend to be smooth, conditions are theoretically favorable for the generation of such small-scale vortices.
In such a scenario, \citet{bib:richard2016} pointed out that the vortex is expected to dissipate rapidly.
As a consequence, the localized accumulation of dust would be a transient phenomenon driven by short-lived gas dynamics, and its overall impact on the long-term formation of a planetary system at this radius would be limited.

\subsection{Upper limit of gas emission associated with the blob}
Despite the high sensitivity of our observations, we did not detect any compact \ce{^13CO} or \ce{C^18O} gas emission associated with the continuum blob.
To infer the local gas environment around the blob, it is crucial to consider the optical depth of these molecular lines.
CO is highly abundant in the upper layers of the disk atmosphere, making its emission typically optically thick.
Previous studies have demonstrated that even the \ce{^13CO} emission can be optically thick in the TW~Hya disk \citep{bib:zhang2017,bib:nomura2021}.
Consequently, our observations likely do not trace the conditions at the disk midplane where the dust blob is embedded, but rather provide an upper limit on the gas column density distribution in the upper atmospheric layers.
Although \ce{C^18O} is more likely to be optically thin in the outer disk, the mass constraint derived from the observational upper limit is more than an order of magnitude higher than that derived from \ce{^13CO}, suggesting that our \ce{C^18O} data may simply lack sufficient sensitivity to detect a compact gas counterpart associated with the blob.

Another factor contributing to the non-detection of molecular line emission is the potential freeze-out of CO molecules near the disk midplane.
The dust continuum blob is physically located near the disk midplane of the PPD.
Previous studies suggest that the local midplane temperature at 52~au is approximately 15--18~K \citep{bib:zhang2017,bib:macias2021,bib:tsukagoshi2022}.
Because the sublimation temperature of CO is roughly 20~K, the bulk of the CO gas in this region is expected to be frozen out onto dust grain surfaces.
Even if an actively forming planet resides at the center of the blob, the thermal impact of the circumplanetary environment would be highly localized.
Theoretical models demonstrate that the region within a CPD where the temperature exceeds the CO sublimation threshold is confined to the vicinity of the planet ($\lesssim$1~au), even for a Jupiter-mass object \citep{bib:zhu2018}.
In such a scenario, any compact gas emission originating from the CO gas within the blob would be heavily suppressed, naturally explaining the absence of an observable gas counterpart.

Despite these observational limitations regarding the midplane,the non-detection of gas in the upper atmospheric layers still provides important constraints on the nature of the embedded structure.
Theoretical hydrodynamic simulations predict that even if a planet is forming deep at the disk midplane, the spiral arms induced by the planet can extend vertically up to the disk atmosphere \citep{bib:zhu2015,bib:zhu2023}.
Our observational results, however, indicate no observable variations in the gas distribution at the emitting layer of these CO isotopologues at the location of the blob.
Because the amplitude of such vertical perturbations strongly depends on the mass of the embedded planet \citep{bib:zhu2015}, the absence of a localized gas signature suggests that if a planet is indeed responsible for the 52~au dust blob, its mass must be sufficiently low so as not to induce significant vertical disturbances in the gas distribution.
This conclusion aligns with the mass constraints derived independently from the dust continuum emission.

\section{Summary}\label{sec:summary}
We present the results of high-resolution ($\sim$1~au) ALMA Band 6 observations of TW~Hya protoplanetary disk, targeting the dust continuum and the \ce{^13CO} and \ce{C^18O} J=2--1 molecular lines.
The primary focus of this study is to investigate the detailed morphology, kinematics, and gas environment of the localized small-scale dust blob previously identified by \citet{bib:tsukagoshi2019a}.
Our main findings and conclusions are summarized as follows:

\begin{itemize}
    \item {By comparing our 2021 continuum data with the archival 2017 data obtained by \citet{bib:tsukagoshi2019a}, we detected the proper motion of the 52~au blob. The measured azimuthal velocity is 3.3$\pm$0.9~km~s$^{-1}$, which is in good agreement with the expected Keplerian velocity at this radius. We found no significant radial migration over the four-year duration of the observations, indicating that the dust structure is robustly co-moving with the disk system.}

    \item {Our high-resolution 2021 continuum image resolved the blob into two distinct peaks separated by approximately 1.7~au along the azimuthal direction. To verify this finding, we applied a sparse-modeling image reconstruction technique to the independent 2017 dataset, successfully confirming the presence of the same double-peaked substructure. This rules out the possibility of an imaging artifact and establishes that the double-peaked morphology is a real physical feature.}

    \item {We found no compact gas emission counterparts associated with the continuum blob in either the \ce{^13CO} or \ce{C^18O} lines. While these observations likely probe the upper atmospheric layers rather than the disk midplane due to high optical depths, the absence of vertical gas perturbations suggests that if an embedded planet is responsible for the blob, its mass must be exceptionally low.}

    \item {We discuss potential physical mechanisms that could produce the observed double-peaked morphology at 52~au: (1) an inclined circumplanetary disk (CPD) with an inner dust cavity or a dusty ring; (2) the roots of macroscopic spiral arms induced by a Neptune-mass planet forming a dense envelope; and (3) alternative scenarios without an actively forming planet, such as the U-turn trajectory of secondary dust due to aerodynamic gas drag, or a purely hydrodynamic, short-lived gas vortex formed by disk instabilities.}
\end{itemize}

The detailed substructure of the 52~au blob in the TW~Hya disk provides an observational testbed for theories of disk-planet interactions and disk instabilities.
To unveil the origin of this double-peaked substructure, future multi-wavelength, high-sensitivity observations, especially at lower frequencies (e.g., ALMA Band 3 or 4, or ngVLA), will be essential to accurately determine the optical depth of the dust component.
Furthermore, deep observations of the gas emission are crucial. Utilizing optically thinner molecular lines that can probe closer to the disk midplane, along with specific molecular tracers sensitive to accretion shocks, such as S-bearing species, could provide direct evidence for the presence of an actively forming planet.
These future observational efforts will determine the physical origin of this localized structure.

\begin{ack}
    We would like to thank the anonymous referee for the constructive comments and suggestions that improved this paper.
    This paper makes use of the following ALMA data: ADS/JAO.ALMA\#2018.1.01173.S, \#2017.1.00520.S, \#2016.1.00842.S, \#2016.1.00229.S, and \#2016.1.01375.S.
    ALMA is a partnership of ESO (representing its member states), NSF (USA)
    and NINS (Japan), together with NRC (Canada), NSTC and ASIAA (Taiwan),
    and KASI (Republic of Korea), in cooperation with the Republic of Chile.
    The Joint ALMA Observatory is operated by ESO, AUI/NRAO and NAOJ. A part
    of the data analysis was carried out on the Multi-wavelength Data
    Analysis System operated by the Astronomy Data Center (ADC), National Astronomical
    Observatory of Japan. This work was supported by JSPS KAKENHI grant No. JP23K20872 and JP24K07097.

    This paper makes use of the following softwares: AnalysisUtilities
    \citep{bib:analysis_utils2023}, Astropy \citep{bib:astropy2022}, the
    Common Astronomy Software Applications \citep[CASA; ][]{bib:casa2022}, matplotlib
    \citep{bib:hunter2007}, Numpy \citep{bib:harris2020}, and PRIISM \citep{bib:priism}.
\end{ack}

\bibliographystyle{aasjournal}
\bibliography{cite}
\end{document}